# Weak ferromagnetism in the square-lattice Heisenberg $J_1 - J_2 - J_2'$ model with easy-axis single-ion anisotropy

Bin-Zhou Mi[a,*], Qiang Gu[b]

[a]School of Science, University of Emergency Management, Beijing 101601, China

[b]Institute of Theoretical Physics and Department of Physics, University of Science and Technology Beijing, Beijing 100083, China

[*]Corresponding author: Email address: mbzfjerry2010@ncist.edu.cn (Bin-Zhou Mi)

## Abstract

Motivated by recent experimental realizations of the Heisenberg model and two-dimensional altermagnetism using ultracold atoms in optical lattices, we systematically investigate the magnetic properties of the square-lattice Heisenberg $J_1 - J_2 - J_2'$ model in the presence of easy-axis single-ion anisotropy $D$. Here, $J_1$ $(> 0)$ is the nearest-neighbor antiferromagnetic exchange parameter, while $J_2$ and $J_2'$ are two distinct next-nearest-neighbor superexchange parameters that alternate on the lattice. We perform numerical calculations of the sublattice magnetization, net magnetization, and critical temperature. Interestingly, the net magnetization is not identically zero at finite temperatures; its absolute value first increases and then decreases with increasing temperature, revealing a temperature-dependent weak ferromagnetism. We find that $J_2 \neq J_2'$ is the primary factor responsible for weak ferromagnetism at finite temperatures, while $D$ provides the necessary background for two-dimensional long-range magnetic ordering. Moreover, for a fixed $J_2 + J_2'$, we observe that a larger $|J_2 - J_2'|$ leads to a greater maximum net magnetization and critical temperature, as well as a broader temperature range for weak ferromagnetism. The difference between $J_2$ and $J_2'$ originates from the distinct superexchange pathways between magnetic atoms, with these pathways mediated by two different types of nonmagnetic atoms. This mechanism, which differs from conventional ones such as the Dzyaloshinskii-Moriya interaction, may be detectable in real materials.

## 1. Introduction

Weak ferromagnetism in antiferromagnets has attracted great interest due to its seemingly anomalous behavior [1-12]. In a conventional antiferromagnet, neighboring spins align antiparallel via superexchange to minimize the exchange energy. However, this perfect compensation (zero net magnetization) can be lifted when inversion symmetry is broken in the crystal structure, giving rise to a relativistic antisymmetric exchange, known as the Dzyaloshinskii–Moriya interaction (DMI) [3, 4]. Its energy term is given by $-\boldsymbol{D}_{ij} \cdot (\boldsymbol{S}_i \times \boldsymbol{S}_j)$, with $\boldsymbol{D}_{ij}$ being the DMI vector between sites $i$ and $j$ [13]. This term energetically favors a mutually orthogonal spin configuration and thus directly competes with the dominant superexchange interaction, which prefers an antiparallel spin arrangement. The resultant competition induces a slight canting of the spins to deviate from the antiparallel alignment, yielding a weak net magnetization.

Weak ferromagnetism, previously known only in conventional antiferromagnets, has now been predicted in altermagnets [14, 15]. For instance, Autieri et al. suggest that a primary mechanism inducing weak ferromagnetism in altermagnets is the staggered DMI, which is ineffective in conventional antiferromagnets [14]. Additionally, another study predicts a new mechanism in which, when DMI is prohibited, the alternating $g$-tensor anisotropy induces weak orbital ferromagnetism in altermagnets [15]. Indeed, altermagnetism, as an emerging magnetic phase, has recently attracted considerable attention in condensed matter physics [16-22]. Like antiferromagnets, altermagnets exhibit a compensated magnetic order with antiparallel spin alignment. Unlike conventional antiferromagnets, however, this compensated magnetic order in altermagnets is always accompanied by spin-split electronic band structures akin to those in ferromagnets. Conceptually, altermagnets can be understood as a redefined subclass of antiferromagnets. For example, the layered compound $La_2O_3Mn_2Se_2$, long regarded as a conventional antiferromagnet [23], has recently been reclassified as an altermagnet [24-26], revealing a spin-split band structure that had previously been overlooked. In other words, altermagnets uniquely integrate the advantages of both ferromagnetic and conventional antiferromagnetic systems, thus holding promise for next-generation spintronics and magnonics.

Very recently, the effective $J_1 - J_2 - J_2'$ model has been proposed to describe the ground-state magnetism of two-dimensional (2D) altermagnets [27-30], including zero net magnetization, quantum fluctuations, and magnon spectrum of the model. Notably, ultracold atoms in optical lattices provide a versatile platform for realizing various magnetic phases and quantum spin models. For instance, Das et al. proposed a scheme for realizing altermagnetism using ultracold fermionic atoms in optical lattices [31]. Ultracold atoms in optical lattices have also been used to simulate the Heisenberg model and probe its dynamics with high precision and control [32, 33]. In addition, Kunimi and Tomita proposed a scheme for realizing the spin-1/2 $J_1$-$J_2$ Heisenberg model and spin-1 Heisenberg model using ultracold Rydberg atoms [34]. These schemes thus offer promising routes for the experimental realization of the 2D $J_1 - J_2 - J_2'$ model.

Nevertheless, a theoretical issue must be addressed when studying the finite-temperature properties of such 2D models. According to the Mermin–Wagner theorem, an isotropic 2D Heisenberg model does not exhibit long-range magnetic order at finite temperatures [35]. Hence, to properly study the finite-temperature magnetism of the 2D $J_1 - J_2 - J_2'$ model, magnetic anisotropy must be included. In fact, 2D van der Waals layered materials with intrinsic magnetism have been experimentally realized [36-40], indicating that magnetic anisotropy is common in real materials. Motivated by these considerations, in this work we theoretically investigate the magnetic properties of the square-lattice Heisenberg $J_1 - J_2 - J_2'$ model with easy-axis single-ion anisotropy. Notably, our calculations reveal temperature-dependent weak ferromagnetism within this model. The remainder of this paper is organized as follows. In Sec. 2, we introduce the square-lattice Heisenberg $J_1 - J_2 - J_2'$ model with easy-axis single-ion anisotropy, and briefly present the main formulas derived using the double-time Green's function (DTGF) method. In Sec. 3, we carefully calculate the sublattice magnetizations, net magnetization, and critical temperature, and present a detailed discussion of their physical implications. Finally, Sec. 4 summarizes the main conclusions and offers an outlook.

## 2. Model and formulas

As illustrated in Fig. 1, the magnetic atoms, represented by solid white and black circles, constitute a square lattice in the *xy* plane, with the magnetic easy-axis along the *z*-direction. The corresponding spin Hamiltonian for the Heisenberg $J_1 - J_2 - J_2'$ model with easy-axis single-ion anisotropy is given as follows:

$$H = J_1 \sum_{\langle ij \rangle} \boldsymbol{S}_i \cdot \boldsymbol{S}_j + J_2 \sum_{\langle\langle ij \rangle\rangle} \boldsymbol{S}_i \cdot \boldsymbol{S}_j + J_2' \sum_{\langle\langle ij \rangle\rangle'} \boldsymbol{S}_i \cdot \boldsymbol{S}_j - D \sum_i (S_i^z)^2. \tag{1}$$

The first three terms in Eq. (1) represent the Heisenberg isotropic exchange energies, where $\langle ij \rangle$ denotes the nearest neighbors, and $\langle\langle ij \rangle\rangle$ and $\langle\langle ij \rangle\rangle'$ denote the two inequivalent, alternating next nearest neighbors in the square lattice (see Fig. 1). Correspondingly, $J_1\ (> 0)$ is the nearest neighbor (nn) antiferromagnetic exchange parameter, while $J_2$ and $J_2'$ are the two distinct next nearest neighbor (nnn) superexchange parameters, respectively. According to the Mermin-Wagner theorem [35], magnetic anisotropy is essential for the spontaneous magnetization of 2D Heisenberg systems at finite temperatures. The last term introduces the single-ion anisotropy, with $D > 0$ favoring an easy-axis along the *z*-direction. This term induces the spontaneous magnetization but is ineffective for $S = 1/2$ systems. We therefore focus on the cases with $S \geq 1$.

The square lattice in Fig. 1 is divided into two sublattices: sublattice a with spin up (white circles) and sublattice b with spin down (black circles). Therefore, the

Hamiltonian in Eq. (1) can also be written as

$$H = J_1 \sum_{\langle \mathrm{a}i,\mathrm{b}j \rangle} \boldsymbol{S}_{\mathrm{a}i} \cdot \boldsymbol{S}_{\mathrm{b}j} + J_2 \sum_{\langle\langle \mathrm{a}i,\mathrm{a}i' \rangle\rangle} \boldsymbol{S}_{\mathrm{a}i} \cdot \boldsymbol{S}_{\mathrm{a}i'} + J_2' \sum_{\langle\langle \mathrm{a}i,\mathrm{a}i'' \rangle\rangle} \boldsymbol{S}_{\mathrm{a}i} \cdot \boldsymbol{S}_{\mathrm{a}i''}$$

$$+ J_2 \sum_{\langle\langle \mathrm{b}j,\mathrm{b}j' \rangle\rangle} \boldsymbol{S}_{\mathrm{b}j} \cdot \boldsymbol{S}_{\mathrm{b}j'} + J_2' \sum_{\langle\langle \mathrm{b}j,\mathrm{b}j'' \rangle\rangle} \boldsymbol{S}_{\mathrm{b}j} \cdot \boldsymbol{S}_{\mathrm{b}j''} - D \sum_{\mathrm{a}i} (S_{\mathrm{a}i}^z)^2$$

$$- D \sum_{\mathrm{b}j} \left(S_{\mathrm{b}j}^z\right)^2. \quad (2)$$

Comparing Eqs. (2) and (1), one finds that all terms except the first in Eq. (1) are divided into two terms corresponding to sublattices a and b, respectively. There are two distinct nnn exchange interactions, $J_2$ and $J_2'$, between spins on the same sublattice, each of which can be positive, negative, or zero. As is well known, when $J_2 = J_2' = 0$ and $D = 0$, the ground state of the system can be regarded as the quantum Néel antiferromagnetic (NAF) state, where the staggered magnetization is reduced by quantum fluctuations, although the exact ground state of the 2D Heisenberg antiferromagnetic model remains unknown. Intuitively, negative $J_2$ or $J_2'$ favors the NAF state. Conversely, positive $J_2$ or $J_2'$ introduces frustration through competition with $J_1$, which tends to suppress the NAF state. In the weakly frustrated regime, the ground state of the present model remains the NAF state.

This paper aims to calculate the sublattice magnetization and net magnetization over the entire temperature range. The DTGF provides a unified and versatile method for studying Heisenberg spin systems and remains valid at all temperatures [41-48]. Nevertheless, the equation of motion method of DTGF generates a series of equation chains containing high-order DTGFs, which require appropriate truncation. In this study, we employ the random phase approximation (RPA) and Anderson-Callen's (ACs) decoupling to truncate the higher-order DTGF. The derivation, though lengthy, is straightforward and follows standard procedures [41-48]. We therefore present only the final formulas required for calculating the sublattice magnetizations, which read as follows:

$$\langle S_\mu^z \rangle = \frac{(\Phi_\mu + 1 + S)\Phi_\mu^{2S+1} - (\Phi_\mu - S)\left(\Phi_\mu + 1\right)^{2S+1}}{\left(\Phi_\mu + 1\right)^{2S+1} - \Phi_\mu^{2S+1}}, (\mu = \mathrm{a}, \mathrm{b}). \quad (3)$$

Where

$$\Phi_\mathrm{a} = \frac{2}{N} \sum_{\boldsymbol{k}} \left\{ \frac{1}{2(\omega_1 - \omega_2)} \left[ (P_{11} - \omega_2) coth\left(\frac{\omega_1}{2k_B T}\right) - (P_{11} - \omega_1) coth\left(\frac{\omega_2}{2k_B T}\right) \right] \right\} - \frac{1}{2}, \quad (4)$$

and

$$\Phi_\mathrm{b} = \frac{2}{N} \sum_{\boldsymbol{k}} \left\{ \frac{1}{2(\omega_1 - \omega_2)} \left[ (P_{22} - \omega_2) coth\left(\frac{\omega_1}{2k_B T}\right) - (P_{22} - \omega_1) coth\left(\frac{\omega_2}{2k_B T}\right) \right] \right\} - \frac{1}{2}. \quad (5)$$

In Eqs. (4) and (5), $k_B$ is the Boltzmann constant, $T$ denotes the temperature of the system. The whole 2D crystal has $N$ magnetic sites, with each sublattice occupying half of them. There are two branches of spin wave spectra written as follows:

$$\omega_1 = \frac{(P_{11} + P_{22}) + \sqrt{(P_{11} - P_{22})^2 + 4P_{12}P_{21}}}{2}, \quad (6)$$

and

$$\omega_2 = \frac{(P_{11} + P_{22}) - \sqrt{(P_{11} - P_{22})^2 + 4P_{12}P_{21}}}{2}. \quad (7)$$

Where

$$P_{11} = -4J_1\langle S_{\mathrm{b}}^z\rangle + [-2J_2 - 2J_2' + J_{2\mathrm{a}}(\boldsymbol{k}) + J_{2\mathrm{a}}'(\boldsymbol{k}) + 2DC_{\mathrm{a}}]\langle S_{\mathrm{a}}^z\rangle, \quad (8)$$

$$P_{22} = -4J_1\langle S_{\mathrm{a}}^z\rangle + [-2J_2 - 2J_2' + J_{2\mathrm{b}}(\boldsymbol{k}) + J_{2\mathrm{b}}'(\boldsymbol{k}) + 2DC_{\mathrm{b}}]\langle S_{\mathrm{b}}^z\rangle, \quad (9)$$

$$P_{12} = J_1(\boldsymbol{k})\langle S_{\mathrm{a}}^z\rangle, \quad (10)$$

$$P_{21} = J_1(\boldsymbol{k})\langle S_{\mathrm{b}}^z\rangle, \quad (11)$$

$$J_1(\boldsymbol{k}) = 2J_1\left[cos(k_x c) + cos(k_y c)\right], \quad (12)$$

$$J_{2\mathrm{a}}(\boldsymbol{k}) = 2J_2 cos(k_x c + k_y c), \quad (13)$$

$$J_{2\mathrm{a}}'(\boldsymbol{k}) = 2J_2' cos(k_x c - k_y c), \quad (14)$$

$$J_{2\mathrm{b}}(\boldsymbol{k}) = 2J_2 cos(k_x c - k_y c), \quad (15)$$

$$J_{2\mathrm{b}}'(\boldsymbol{k}) = 2J_2' cos(k_x c + k_y c), \quad (16)$$

$$C_\mu = 1 - \frac{1}{2S^2}\left[S(S+1) - \langle S_\mu^z S_\mu^z\rangle\right], (\mu = \mathrm{a}, \mathrm{b}), \quad (17)$$

and

$$\langle S_\mu^z S_\mu^z\rangle = S(S+1) - \left(1 + 2\Phi_\mu\right)\langle S_\mu^z\rangle, (\mu = \mathrm{a}, \mathrm{b}). \quad (18)$$

For convenience, the two sublattice magnetizations are denoted as $m_{\mathrm{a}} = \langle S_{\mathrm{a}}^z\rangle$ and $m_{\mathrm{b}} = \langle S_{\mathrm{b}}^z\rangle$. The net magnetization $M$ is the sum of these two quantities:

$$M = m_{\mathrm{a}} + m_{\mathrm{b}} = \langle S_{\mathrm{a}}^z\rangle + \langle S_{\mathrm{b}}^z\rangle. \quad (19)$$

In Eqs. (12)-(16), $\boldsymbol{k}$ is the wave vector, with components $k_x$ and $k_y$ along the $x$ and $y$ directions, respectively, and $c$ is the distance between nn sites. Eqs. (3)-(19) are the transcendental equations from which the sublattice magnetizations and net magnetization can be calculated as functions of temperature, spin quantum number, exchange parameters, and single-ion anisotropy strength.

## 3. Results and discussions

In numerical calculations, the nn antiferromagnetic exchange parameter $J_1 = 1$ is fixed as the energy unit of the system, and $J_2$, $J_2'$, and $D$ are varied. For convenience, the Boltzmann constant is set to $k_B = 1$. Consequently, all physical quantities, including the sublattice magnetization, net magnetization, and critical temperature, are dimensionless. As mentioned in Sec. 2, $J_2$ and $J_2'$ may take positive, negative, or zero values. Therefore, we discuss three representative parameter sets characterized by $J_2 + J_2' =$ 0.8, -0.8, and 0.

Figures 2(a)-2(d) plot the sublattice magnetizations $m_{\mathrm{a}}$ and $-m_{\mathrm{b}}$, as well as the net magnetization $M$, as functions of temperature $T$ for four sets of parameters, with $S = 1$ and $D = 0.08$. The four parameters sets are: (a) $J_2 = 0$, $J_2' = 0.8$; (b) $J_2 = 0.8$, $J_2' = 0$; (c) $J_2 = 0.2$, $J_2' = 0.6$; and (d) $J_2 = 0.6$, $J_2' = 0.2$. It is evident that the sum of $J_2$ and $J_2'$ remains constant, i.e., $J_2 + J_2' = 0.8$. For convenience, the sublattice magnetizations at zero temperature ($T = 0$ K) are denoted as $m_{\mathrm{a}}(0)$ and $m_{\mathrm{b}}(0)$, respectively. It can be seen from Figs. 2(a)-2(d) that both $m_{\mathrm{a}}(0)$ and $-m_{\mathrm{b}}(0)$ are equal to each other and smaller than the saturation value $S = 1$. Consequently, the net magnetization $M$ is zero. At very low temperatures, the sublattice magnetizations $m_{\mathrm{a}}$ and $-m_{\mathrm{b}}$ remain almost unchanged, exhibiting flat curves, and the net magnetization $M$ remains zero. This is because at sufficiently low temperatures, only the infinite-wavelength (zero-wave-vector) spin-wave modes are populated. These long-wavelength modes are insensitive to differences in the neighboring exchange couplings of the a and b sublattices, and thus the two sublattice magnetizations have the same magnitude. The sublattice magnetizations $m_{\mathrm{a}}$ and $-m_{\mathrm{b}}$ decrease at different rates as the temperature further increases, and the corresponding net magnetization becomes nonzero. This is because spin waves with finite wavelengths are thermally excited. When a spin wave propagates along a given direction, it can sense the difference in local environments between the a and b sublattices. As illustrated in Fig. 1, the nnn bonds of an a*i*-site are rotated by 90° relative to those of a b*j*-site. Consequently, the sublattice magnetizations are no longer perfectly compensated ($m_{\mathrm{a}} \neq -m_{\mathrm{b}}$), yielding a small nonzero net magnetization, which is the characteristic of weak ferromagnetism. For example, Figs. 2(a) and 2(c) show that the sublattice magnetization $m_{\mathrm{a}}$ initially decreases more slowly than $-m_{\mathrm{b}}$ does as temperature rises. However, beyond a certain temperature, the former decreases faster than the latter. The temperature at which both $m_{\mathrm{a}}$ and $-m_{\mathrm{b}}$ vanish is the critical temperature $T_{\mathrm{C}}$. When the temperature approaches $T_{\mathrm{C}}$, the decreasing rates of $m_{\mathrm{a}}$ and $-m_{\mathrm{b}}$ tend to be the same. Thereupon, the difference between $m_{\mathrm{a}}$ and $-m_{\mathrm{b}}$ first increases and then decreases with rising temperature. Correspondingly, the net magnetization first rises from zero to a maximum and then falls back to zero, forming a dome-shaped curve.

For clarity, the enlarged curves of $M$ versus $T$ are presented in the corresponding insets of Figs. 2(a)-2(d). Obviously, the net magnetization is not identically zero at finite temperatures; its absolute value first increases and then decreases with increasing temperature, reaching a maximum value denoted $M_{\mathrm{max}}$. The parameters $J_2$ and $J_2'$ in Fig. 2(b) are obtained by swapping the parameters $J_2$ and $J_2'$ in Fig. 2(a), and the parameters in Fig. 2(c) and 2(d) are also swapped with each other. The interchange of $J_2$ and $J_2'$ results in the interchange of sublattice magnetizations $m_{\mathrm{a}}$ and $-m_{\mathrm{b}}$. Hence, the two corresponding net magnetizations have equal magnitudes but opposite signs, as seen in the insets of Figs. 2(a) and 2(b), as well as the insets of Figs. 2(c) and 2(d).

For further analysis, Table 1 lists the maximum net magnetization $M_{\mathrm{max}}$, the ground-state sublattice magnetization $m_{\mathrm{a}}(0)$ ( $=|m_{\mathrm{b}}(0)|$ ), and the critical temperature $T_{\mathrm{C}}$ for nine sets of $J_2$ and $J_2'$, including the four sets shown in Figs. 2(a)-2(d). As seen in Table 1, the ground-state sublattice magnetizations are equal for all cases and depend solely on the sum $J_2+J_2'$. For $J_2+J_2'=0.8$, the larger the $|J_2-J_2'|$, the greater the values of $M_{\mathrm{max}}$ and $T_{\mathrm{C}}$, and the broader the temperature range for weak ferromagnetism, as can be seen by comparing Figs. 2(a) with 2(c), or Figs. 2(b) with 2(d). The symmetric case $J_2=J_2'=0.4$, corresponding to ordinary NAF state, yields the minimum values of both $M_{\mathrm{max}}$ and $T_{\mathrm{C}}$. Moreover, Table 1 confirms that both $M_{\mathrm{max}}$ and $T_{\mathrm{C}}$ are invariant under the interchange of $J_2$ and $J_2'$ ($J_2 \leftrightarrow J_2'$).

In particular, for the symmetric case with $J_2=J_2'=0.4$ in Table 1, the net magnetization vanishes at all temperatures, which indicates that the asymmetry $J_2 \neq J_2'$ may be the primary factor in the emergence of weak ferromagnetism at finite temperatures. According to Eqs. (4)-(7), we can obtain the relationship $\Phi_{\mathrm{a}}+\Phi_{\mathrm{b}}=\frac{2}{N}\sum_{\boldsymbol{k}}\left\{\frac{1}{2}\left[\coth\left(\frac{\omega_1}{2k_{\mathrm{B}}T}\right)+\coth\left(\frac{\omega_2}{2k_{\mathrm{B}}T}\right)\right]\right\}-1$. If $J_2=J_2'$, then $\omega_1=-\omega_2$, so $\Phi_{\mathrm{a}}+\Phi_{\mathrm{b}}=-1$, which corresponds to the case of a conventional antiferromagnet [47, 48]. In contrast, for $J_2 \neq J_2'$, whether $\Phi_{\mathrm{a}}+\Phi_{\mathrm{b}}$ remains equal to -1 requires further examination. In fact, as long as $J_2 \neq J_2'$, we have $\omega_1 \neq -\omega_2$, which means that the two spin waves are no longer degenerate. From the relationship $\Phi_{\mathrm{a}}+\Phi_{\mathrm{b}}=\frac{2}{N}\sum_{\boldsymbol{k}}\left\{\frac{1}{2}\left[\coth\left(\frac{\omega_1}{2k_{\mathrm{B}}T}\right)+\coth\left(\frac{\omega_2}{2k_{\mathrm{B}}T}\right)\right]\right\}-1$, it can be concluded that at zero and very low temperatures, i.e. $T=0$ K and $T\to 0$ K, $\Phi_{\mathrm{a}}+\Phi_{\mathrm{b}}$ remains equal to -1. Correspondingly, the two sublattice magnetizations exactly compensate each other, resulting in zero net magnetization. As the temperature further increases, this perfect compensation is broken because $\Phi_{\mathrm{a}}+\Phi_{\mathrm{b}}\neq -1$. To further investigate this mechanism of weak ferromagnetism, we also performed calculations for the other two cases with $J_2+J_2'=$ -0.8 and 0, as presented below.

Figures 3(a)-3(d) depict the sublattice magnetizations $m_{\mathrm{a}}$ and $-m_{\mathrm{b}}$, along with the net magnetization $M$, as functions of temperature $T$ for four parameter sets with $J_2 + J_2' = -0.8$ at $S = 1$ and $D = 0.08$. To clearly show the variation of the net magnetization, magnified plots of $M$ versus $T$ are also presented respectively in the insets of Figs. 3(a)-3(d). Correspondingly, Table 2 lists the values of $M_{\max}$, $m_{\mathrm{a}}(0)$, and $T_{\mathrm{C}}$ for nine parameter sets $(J_2, J_2')$ satisfying $J_2 + J_2' = -0.8$, including the four sets shown in Figs. 3(a)-3(d). We also find that the net magnetization becomes nonzero at finite temperatures; its absolute value first increases and then decreases with rising temperature, indicating weak ferromagnetism whenever $J_2 \neq J_2'$. Compared with the parameter sets in Table 1, the negative $J_2$ and $J_2'$ values in Table 2 favor the NAF state. Consequently, $m_{\mathrm{a}}(0)$ and $T_{\mathrm{C}}$ are higher than those in Table 1. However, the differences in the corresponding $M_{\max}$ values between the two tables are relatively small.

To proceed, Figs. 4(a)-4(d) illustrate the sublattice magnetizations $m_{\mathrm{a}}$ and $-m_{\mathrm{b}}$, as well as the net magnetization $M$, versus temperature $T$ for four parameter sets with $J_2 + J_2' = 0$ at $S = 1$ and $D = 0.08$. For clarity, magnified plots of $M$ versus $T$ are depicted in the insets of Figs. 4(a)-4(d). Correspondingly, Table 3 lists the values of $M_{\max}$, $m_{\mathrm{a}}(0)$, and $T_{\mathrm{C}}$ for nine parameter sets $(J_2, J_2')$ satisfying $J_2 + J_2' = 0$, including the four sets shown in Figs. 4(a)-4(d). We also observe that the absolute value of net magnetization first increases from zero to a maximum and then decreases back to zero, revealing a temperature-dependent weak ferromagnetism whenever $J_2 \neq J_2'$. In these parameter sets $(J_2, J_2')$, a positive $J_2$ or $J_2'$ introduces frustration, whereas a negative one does not. Thus, among the three cases ($J_2 + J_2' =$ 0.8, -0.8, and 0) considered, the case with $J_2 + J_2' = 0$ exhibits the largest $|J_2 - J_2'|$. Comparing Table 3 with Tables 1 and 2, and Fig. 4 with Figs. 2 and 3, we observe that a larger $|J_2 - J_2'|$ leads to a correspondingly greater $M_{\max}$.

In the following, the parameters $J_2$ and $J_2'$ are fixed, while $D$ and $S$ are varied to calculate $m_{\mathrm{a}}(0)$ and $M_{\max}$. Figures 5(a) and 5(b) plot, respectively, the ground-state sublattice magnetization $m_{\mathrm{a}}(0)$ and the maximum net magnetization $M_{\max}$ versus single-ion anisotropy strength $D$ at $S = 1$, $J_2 = -1.2$, and $J_2' = 0.8$. Figure 5(a) shows that $m_{\mathrm{a}}(0)$ increases with increasing $D$, due to the suppression of quantum spin fluctuations by single-ion anisotropy. Similarly, Fig. 5(b) shows that $M_{\max}$ also increases with increasing $D$. Notably, $M_{\max}$ remains nonzero as $D$ approaches zero. Consequently, we conclude that $J_2 \neq J_2'$ is the primary factor responsible for weak ferromagnetism at finite temperatures, while $D$ serves only as the background for the formation of 2D long-range magnetic ordering. It is worth noting that, unlike the DMI mechanism, the difference in these two exchange parameters arises from distinct superexchange pathways between magnetic atoms, which are mediated by two different types of nonmagnetic atoms [24]. This suggests that such weak ferromagnetism may be observable in 2D materials, such as monolayer $La_2O_3Mn_2Se_2$, which could be synthesized in the future. Alternatively, this mechanism could be simulated using ultracold atoms in optical lattices [31-34].

Finally, Figs. 6(a) and 6(b) depict, respectively, the normalized ground-state sublattice magnetization $m_{\mathrm{a}}(0)/S$ and the normalized maximum net magnetization $M_{\mathrm{max}}/S$ versus spin quantum number $S$ at $J_2 = -1.2$, $J_2' = 0.8$, and $D = 0.02$. Figure 6(a) demonstrates that $m_{\mathrm{a}}(0)/S$ increases with rising $S$, because large spin quantum numbers suppress quantum fluctuations and drive the system toward the classical limit. Figure 6(b) indicates that $M_{\mathrm{max}}/S$ initially increases rapidly with $S$, reaches a maximum at $S = 6$, and then slowly decreases as $S$ further increases. A possible explanation for this unusual behavior is that the temperature corresponding to $M_{\mathrm{max}}$ increases with $S$; when $S > 6$, thermal fluctuations dominate over quantum fluctuations and suppress the net magnetization of the system.

## 4. Concluding remarks

In summary, we have theoretically investigated the magnetic properties of the square-lattice Heisenberg $J_1 - J_2 - J_2'$ model with easy-axis single-ion anisotropy $D$ by employing the DTGF method within the RPA and ACs decoupling. We calculated the sublattice magnetization, net magnetization, and critical temperature for three representative parameter sets with $J_2 + J_2' =$ 0.8, -0.8, and 0. Our results demonstrate that $J_2 \neq J_2'$ is the primary origin of weak ferromagnetism at finite temperatures, while $D$ serves merely as the background for the formation of 2D long-range magnetic ordering.

Microscopically, at zero and very low temperatures, only spin-wave modes with infinite-wavelength (zero-wave-vector) are populated. The two sublattice magnetizations exactly compensate each other, resulting in zero net magnetization. Above a certain finite temperature, however, the thermal excitation of spin waves with finite wavelengths lifts this compensation, giving rise to a small net magnetization, i.e., the temperature-dependent weak ferromagnetism. The absolute value of net magnetization first increases and then decreases with increasing temperature, reaching a maximum value $M_{\mathrm{max}}$ at a finite intermediate temperature.

We have observed that for a fixed $J_2 + J_2'$, the larger the $|J_2 - J_2'|$, the greater the values of $M_{\mathrm{max}}$ and $T_{\mathrm{C}}$, and the broader the temperature range for weak ferromagnetism. When $J_2 = J_2'$, the net magnetization vanishes at all temperatures, and $T_C$ reaches its minimum value. However, the ground-state sublattice magnetization $m_a(0)$ is independent of the difference between $J_2$ and $J_2'$, depending only on the sum $J_2 + J_2'$. Additionally, the results for $M_{max}$ and $T_C$ are symmetric under the interchange of $J_2$ and $J_2'$.

We have also calculated the effects of $D$ and $S$ on $M_{max}$ and $m_a(0)$. The numerical results show that both $M_{max}$ and $m_a(0)$ increase monotonically with $D$. It is worth noting that $M_{max}$ remains nonzero even as $D$ approaches zero, confirming that single-ion anisotropy only provides a background for weak ferromagnetism, as described above. The normalized ground-state sublattice magnetization $m_a(0)/S$ increases with the increase of $S$. In contrast, the normalized maximum net magnetization $M_{max}/S$ exhibits an unusual spin dependence: it first increases rapidly with $S$, reaches a maximum at $S = 6$, and then slowly decreases. A possible explanation is that the temperature at which $M_{max}$ appears increases with $S$; when $S > 6$, thermal fluctuations dominate over quantum fluctuations and suppress the net magnetization of the system. Finally, these theoretical results are expected to stimulate experimental studies of weak ferromagnetism in monolayer magnets as well as in ultracold atoms.

## Acknowledgements

Bin-Zhou Mi thanks Prof. Huai-Yu Wang for his helpful discussions. This study was supported by the Natural Science Foundation of Hebei Province of China (Grant No. F2020508001).

## Figure captions:

**Fig. 1.** The magnetic atoms, shown as solid white and black circles, constitute a square lattice in the *xy* plane, with the magnetic easy-axis along the *z*-direction. The parameter $J_1$ denotes the nn antiferromagnetic exchange interaction, while $J_2$ and $J_2'$ are the two distinct and alternating nnn superexchange parameters. There are two sublattices: sublattice a with spin up (white circles) and sublattice b with spin down (black circles). The distance between the nn magnetic atoms is *c*.

**Fig. 2.** (Color online) Temperature dependence of magnetizations, including sublattice magnetization and net magnetization, for $S = 1$ and $D = 0.08$ with $J_2 + J_2' = 0.8$ : (a) $J_2 = 0, J_2' = 0.8$; (b) $J_2 = 0.8, J_2' = 0$; (c) $J_2 = 0.2, J_2' = 0.6$; (d) $J_2 = 0.6, J_2' = 0.2$. Insets: enlarged $M$ versus $T$ curves.

**Fig. 3.** (Color online) Temperature dependence of magnetizations, including sublattice magnetization and net magnetization, for $S = 1$ and $D = 0.08$ with $J_2 + J_2' = -0.8$ : (a) $J_2 = -0.8, J_2' = 0$; (b) $J_2 = 0, J_2' = -0.8$; (c) $J_2 = -0.6, J_2' = -0.2$; (d) $J_2 = -0.2, J_2' = -0.6$. Insets: enlarged $M$ versus $T$ curves.

**Fig. 4.** (Color online) Temperature dependence of magnetizations, including sublattice magnetization and net magnetization, for $S = 1$ and $D = 0.08$ with $J_2 + J_2' = 0$ : (a) $J_2 = -0.8, J_2' = 0.8$; (b) $J_2 = 0.8, J_2' = -0.8$; (c) $J_2 = -0.4, J_2' = 0.4$; (d) $J_2 = 0.4, J_2' = -0.4$. Insets: enlarged $M$ versus $T$ curves.

**Fig. 5.** The variation of (a) $m_{\mathrm{a}}(0)$ and (b) $M_{\max}$ with $D$ for $S = 1$, $J_2 = -1.2$, and $J_2' = 0.8$. The dotted lines are guides to the eye.

**Fig. 6.** (a) $m_{\mathrm{a}}(0)/S$ and (b) $M_{\max}/S$ as a function of $S$ for $J_2 = -1.2$, $J_2' = 0.8$, and $D = 0.02$. The dotted lines are guides to the eye.

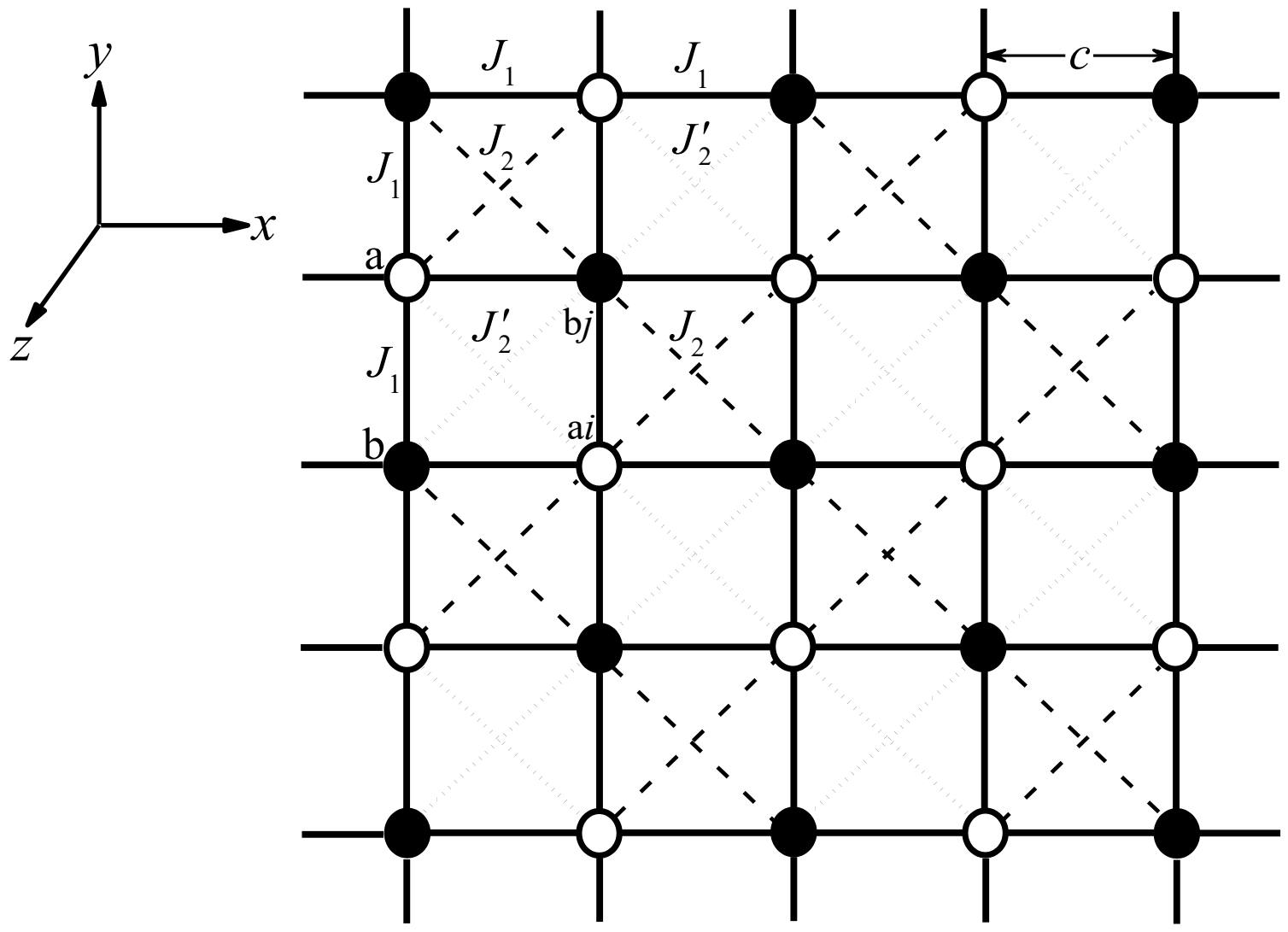

**Fig. 1.** The magnetic atoms, shown as solid white and black circles, constitute a square lattice in the *xy* plane, with the magnetic easy-axis along the *z*-direction. The parameter $J_1$ denotes the nn antiferromagnetic exchange interaction, while $J_2$ and $J_2'$ are the two distinct and alternating nnn superexchange parameters. There are two sublattices: sublattice a with spin up (white circles) and sublattice b with spin down (black circles). The distance between the nn magnetic atoms is *c*.

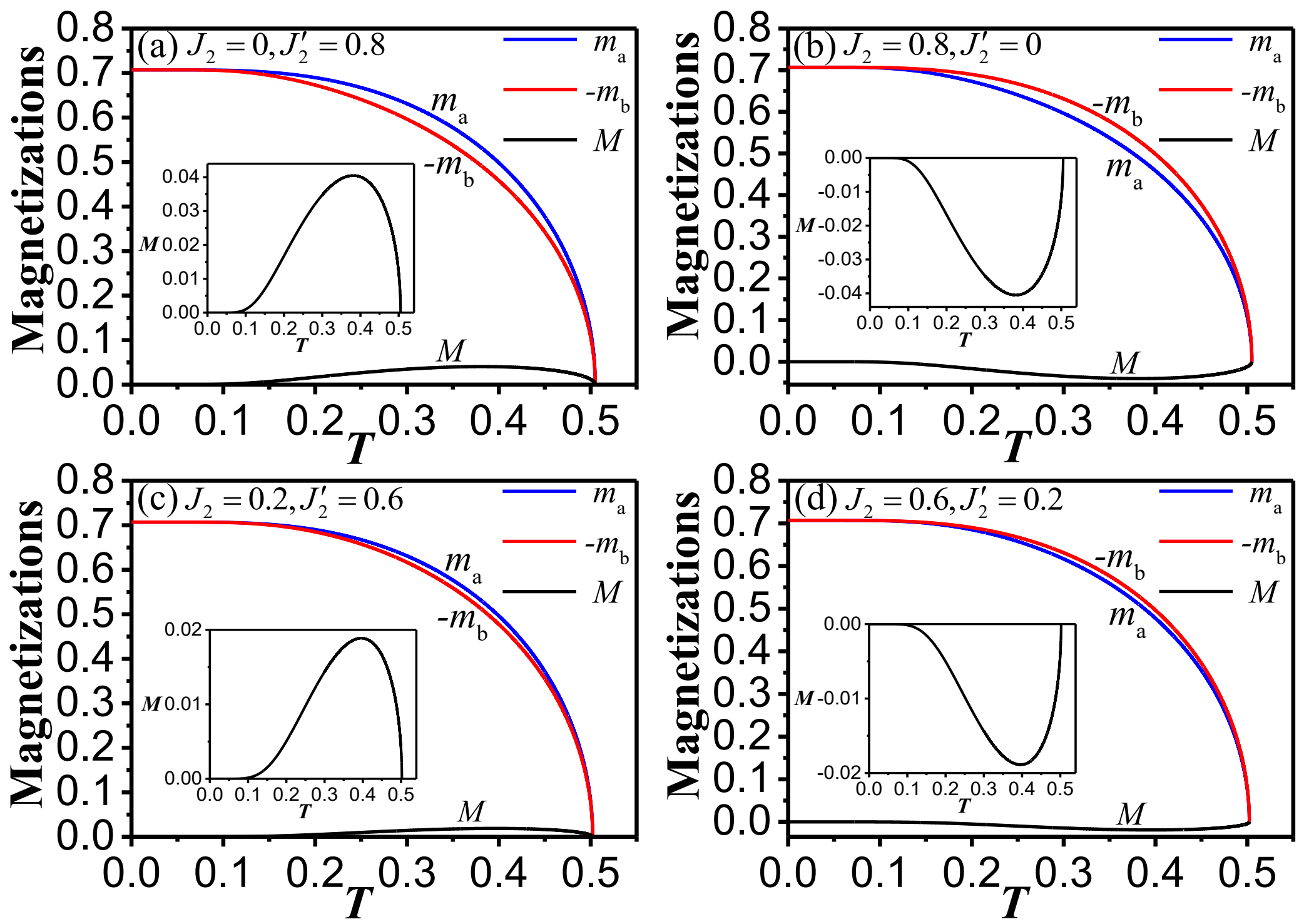

**Fig. 2.** (Color online) Temperature dependence of magnetizations, including sublattice magnetization and net magnetization, for $S = 1$ and $D = 0.08$ with $J_2 + J_2' = 0.8$ : (a) $J_2 = 0, J_2' = 0.8$; (b) $J_2 = 0.8, J_2' = 0$; (c) $J_2 = 0.2, J_2' = 0.6$; (d) $J_2 = 0.6, J_2' = 0.2$. Insets: enlarged $M$ versus $T$ curves.

**Table 1.** Maximum net magnetization, ground-state sublattice magnetization, and critical temperature are listed for several sets of parameters with $J_2 + J_2' = 0.8$, $S = 1$ and $D = 0.08$.

| $J_2$ | $J_2'$ | $M_{\max}$ | $m_a(0) = \lvert m_b(0)\rvert$ | $T_C$ |
|---|---|---|---|---|
| 0.0 | 0.8 | 0.040503 | 0.707052 | 0.504948 |
| 0.1 | 0.7 | 0.029139 | 0.707052 | 0.503585 |
| 0.2 | 0.6 | 0.018879 | 0.707052 | 0.502135 |
| 0.3 | 0.5 | 0.009283 | 0.707052 | 0.501098 |
| 0.4 | 0.4 | 0 (Always zero) | 0.707052 | 0.500727 |
| 0.5 | 0.3 | $\lvert-0.009283\rvert$ | 0.707052 | 0.501098 |
| 0.6 | 0.2 | $\lvert-0.018879\rvert$ | 0.707052 | 0.502135 |
| 0.7 | 0.1 | $\lvert-0.029139\rvert$ | 0.707052 | 0.503585 |
| 0.8 | 0.0 | $\lvert-0.040503\rvert$ | 0.707052 | 0.504948 |

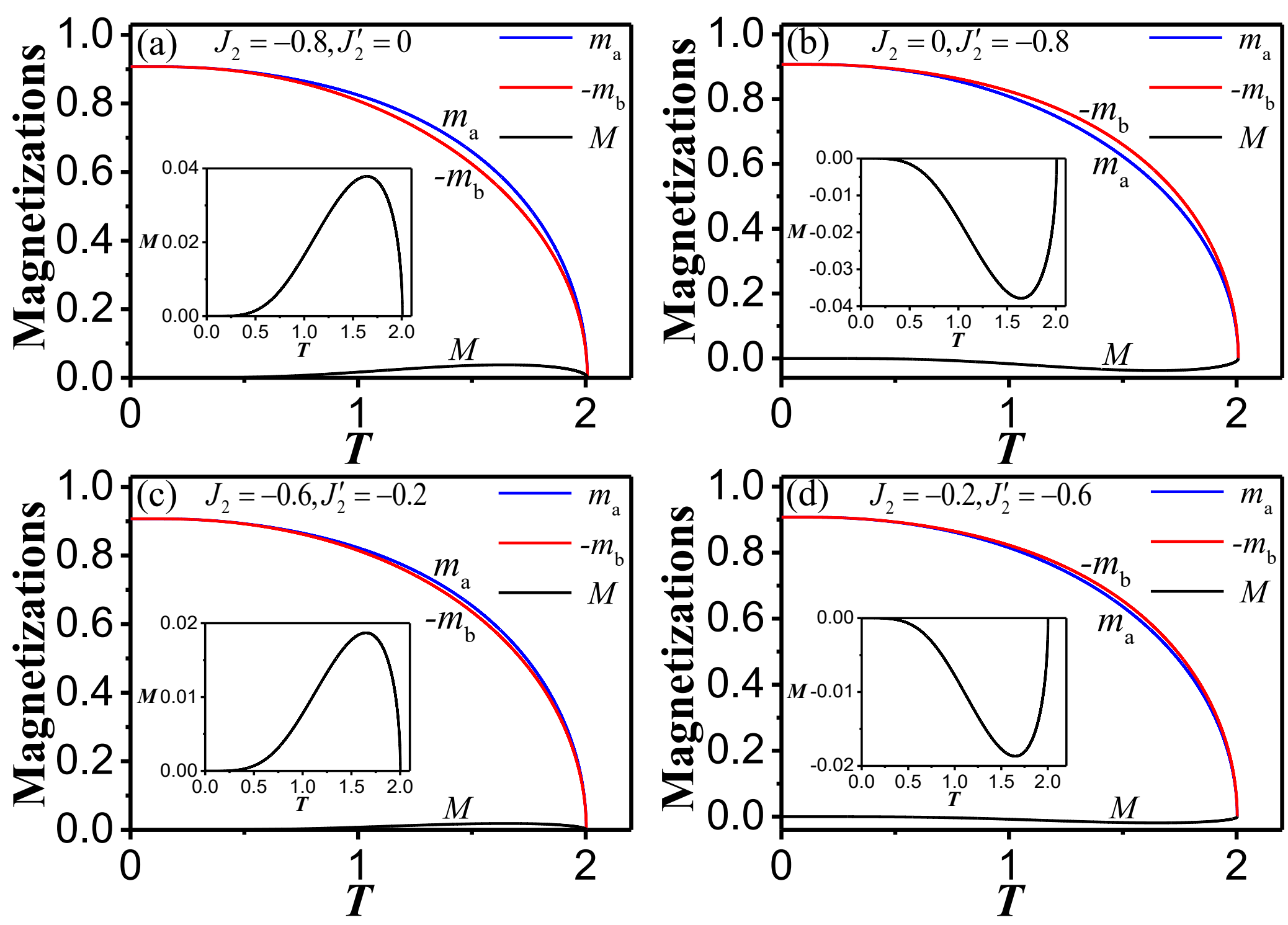

**Fig. 3.** (Color online) Temperature dependence of magnetizations, including sublattice magnetization and net magnetization, for $S=1$ and $D=0.08$ with $J_2+J_2'=-0.8$ : (a) $J_2=-0.8, J_2'=0$; (b) $J_2=0, J_2'=-0.8$; (c) $J_2=-0.6, J_2'=-0.2$; (d) $J_2=-0.2, J_2'=-0.6$. Insets: enlarged $M$ versus $T$ curves.

**Table 2.** Maximum net magnetization, ground-state sublattice magnetization, and critical temperature are listed for several sets of parameters with $J_2+J_2'=-0.8$, $S=1$ and $D=0.08$.

| $J_2$ | $J_2'$ | $M_{\text{max}}$ | $m_a(0)=\lvert m_b(0)\rvert$ | $T_C$ |
|---|---|---|---|---|
| -0.8 | 0.0 | 0.037859 | 0.907262 | 2.007355 |
| -0.7 | -0.1 | 0.028183 | 0.907262 | 2.004556 |
| -0.6 | -0.2 | 0.018689 | 0.907262 | 2.002558 |
| -0.5 | -0.3 | 0.009315 | 0.907262 | 2.001358 |
| -0.4 | -0.4 | 0 (Always zero) | 0.907262 | 2.000961 |
| -0.3 | -0.5 | $\lvert-0.009315\rvert$ | 0.907262 | 2.001358 |
| -0.2 | -0.6 | $\lvert-0.018689\rvert$ | 0.907262 | 2.002558 |
| -0.1 | -0.7 | $\lvert-0.028183\rvert$ | 0.907262 | 2.004556 |
| 0.0 | -0.8 | $\lvert-0.037859\rvert$ | 0.907262 | 2.007355 |

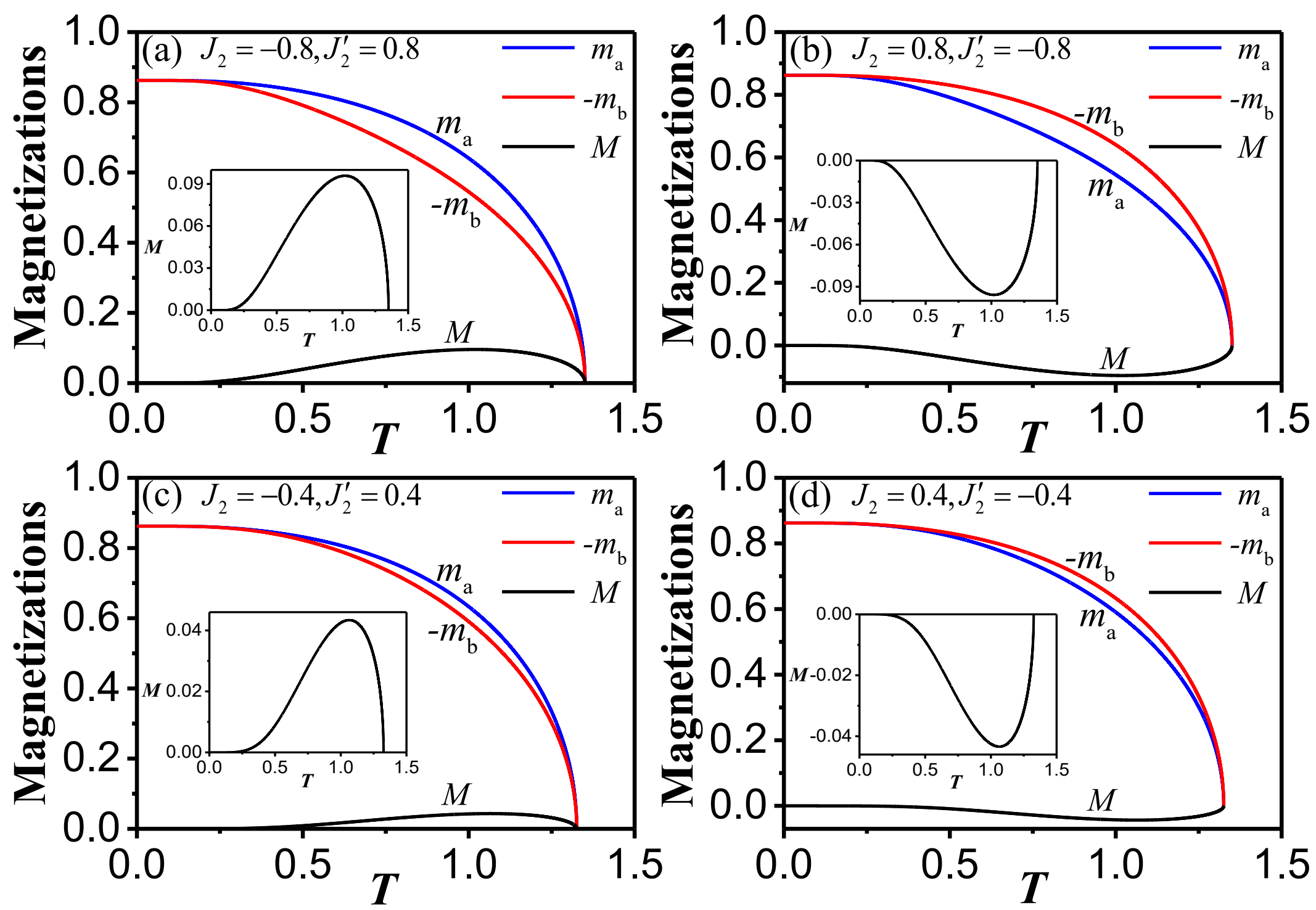

**Fig. 4.** (Color online) Temperature dependence of magnetizations, including sublattice magnetization and net magnetization, for $S = 1$ and $D = 0.08$ with $J_2 + J_2' = 0$ : (a) $J_2 = -0.8, J_2' = 0.8$; (b) $J_2 = 0.8, J_2' = -0.8$; (c) $J_2 = -0.4, J_2' = 0.4$; (d) $J_2 = 0.4, J_2' = -0.4$. Insets: enlarged $M$ versus $T$ curves.

**Table 3.** Maximum net magnetization, ground-state sublattice magnetization, and critical temperature are listed for several sets of parameters with $J_2 + J_2' = 0$, $S = 1$ and $D = 0.08$.

| $J_2$ | $J_2'$ | $M_{\text{max}}$ | $m_a(0) = \lvert m_b(0)\rvert$ | $T_C$ |
|---|---|---|---|---|
| -0.8 | 0.8 | 0.095825 | 0.862228 | 1.350645 |
| -0.6 | 0.6 | 0.067775 | 0.862228 | 1.336094 |
| -0.4 | 0.4 | 0.043397 | 0.862228 | 1.325525 |
| -0.2 | 0.2 | 0.021192 | 0.862228 | 1.319122 |
| 0.0 | 0.0 | 0 (Always zero) | 0.862228 | 1.316980 |
| 0.2 | -0.2 | $\lvert -0.021192\rvert$ | 0.862228 | 1.319122 |
| 0.4 | -0.4 | $\lvert -0.043397\rvert$ | 0.862228 | 1.325525 |
| 0.6 | -0.6 | $\lvert -0.067775\rvert$ | 0.862228 | 1.336094 |
| 0.8 | -0.8 | $\lvert -0.095825\rvert$ | 0.862228 | 1.350645 |

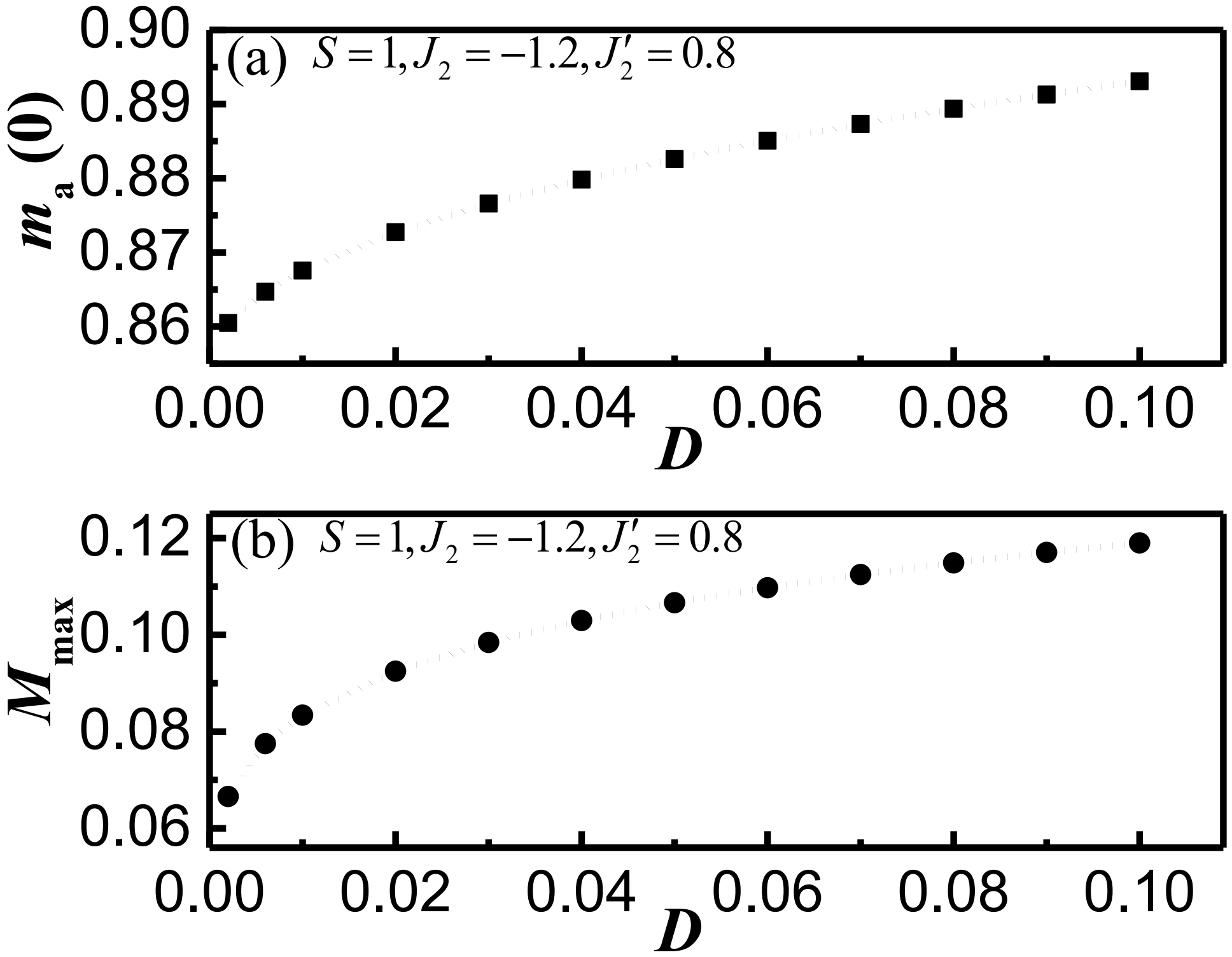

**Fig. 5.** The variation of (a) $m_a(0)$ and (b) $M_{\max}$ with $D$ for $S = 1$, $J_2 = -1.2$, and $J_2' = 0.8$. The dotted lines are guides to the eye.

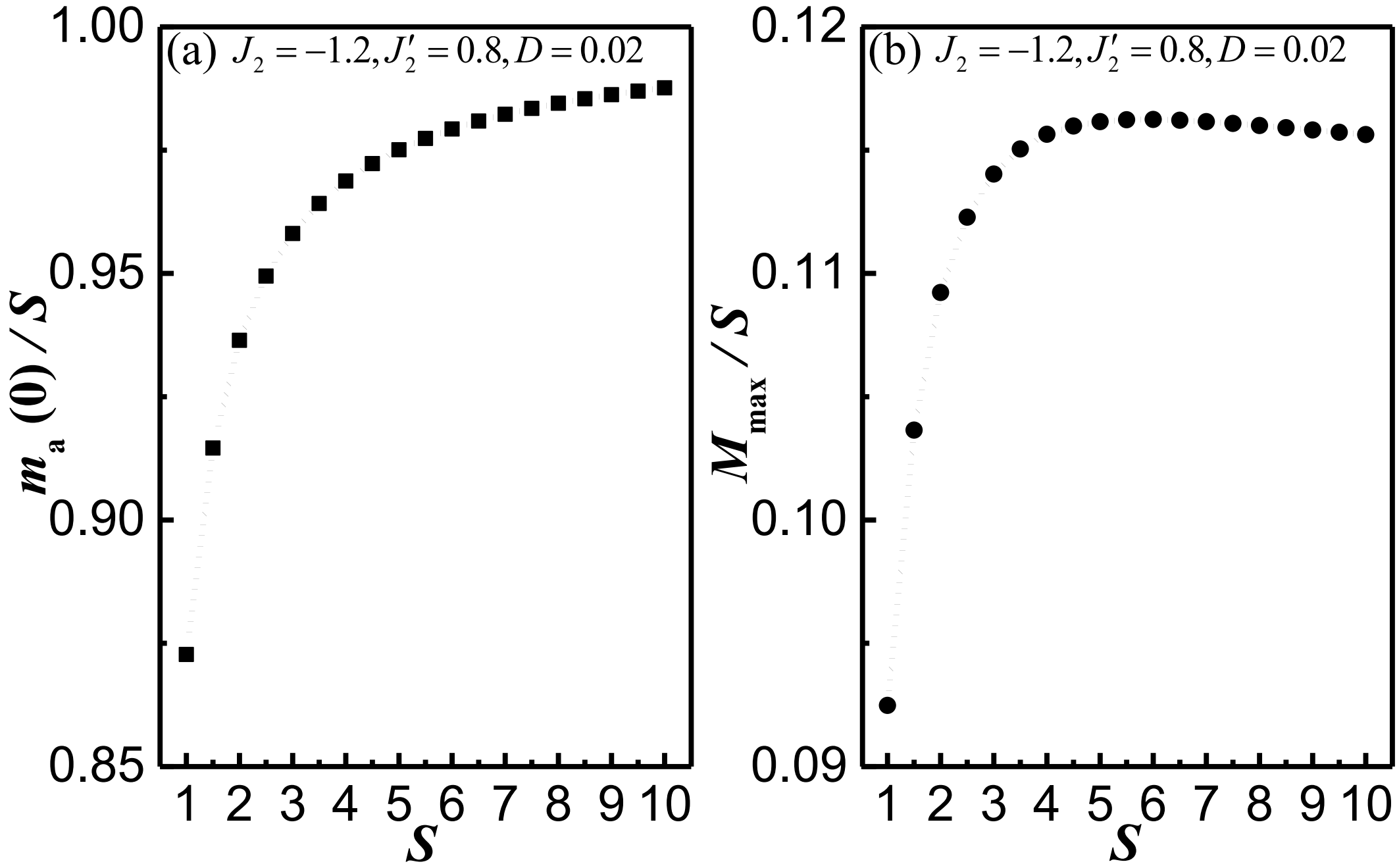

**Fig. 6.** (a) $m_a(0)/S$ and (b) $M_{\max}/S$ as a function of $S$ for $J_2 = -1.2$, $J_2' = 0.8$, and $D = 0.02$. The dotted lines are guides to the eye.